\documentclass[prd,twocolumn]{revtex4}
\usepackage{graphicx}

\def\be{\begin{equation}}
\def\ee{\end{equation}}

\begin{document}
\title{Phenomenological Gaussian screening
in the nonextensive statistics approach to fully developed
turbulence}
 \author{A.K. Aringazin}
 \email{aringazin@mail.kz}
 \affiliation{Department of Theoretical
Physics, Institute for Basic Research, Eurasian National
University, Astana 473021 Kazakstan}
 \author{M.I. Mazhitov}
 \email{mmi@emu.kz}
 \affiliation{Department of Theoretical Physics, Institute for
Basic Research, Eurasian National University, Astana 473021
Kazakstan}

\date{30 December 2002}

\begin{abstract}
We propose a simple phenomenological modification, a Gaussian
screening, of the probability distribution function which was
obtained by Beck to explain experimentally measured distribution
from fully developed fluid turbulence, within the framework of
nonextensive statistical mechanics. The modified distribution
provides a good fit to new experimental results on the
acceleration of fluid particles reported by Crawford, Mordant, and
Bodenschatz. Theoretical foundations of such a modification
requires a separate study.
\end{abstract}
\maketitle

Recently, Beck \cite{Beck} have studied application of Tsallis
formalism \cite{Tsallis}  to turbulent flows and achieved a good
agreement with experimental measurements \cite{Bodenschatz}.
Remarkably, no fitting parameters have been used by Beck to
reproduce histogram of the acceleration $a$ of a test particle
advected by the turbulent flow, while the stretch exponential fit
requires three free parameters for this purpose,
\be\label{1}
P(a) = C
\exp\left[a^2/(1+\left|\frac{a\beta}{\sigma}\right|^\gamma\sigma^2)\right],
\ee
where $\beta= 0.513$, $\gamma= 1.600$, $\sigma= 0.563$ and $C$ is
a normalization constant.

The main idea underlying generalized statistical mechanics
approach to turbulence is to introduce fluctuation of temperature
or fluctuation of energy dissipation \cite{Beck2, Wilk} described
by gamma distribution.

\begin{figure}[tbp!]
\begin{center}
\includegraphics[width=0.4\textwidth]{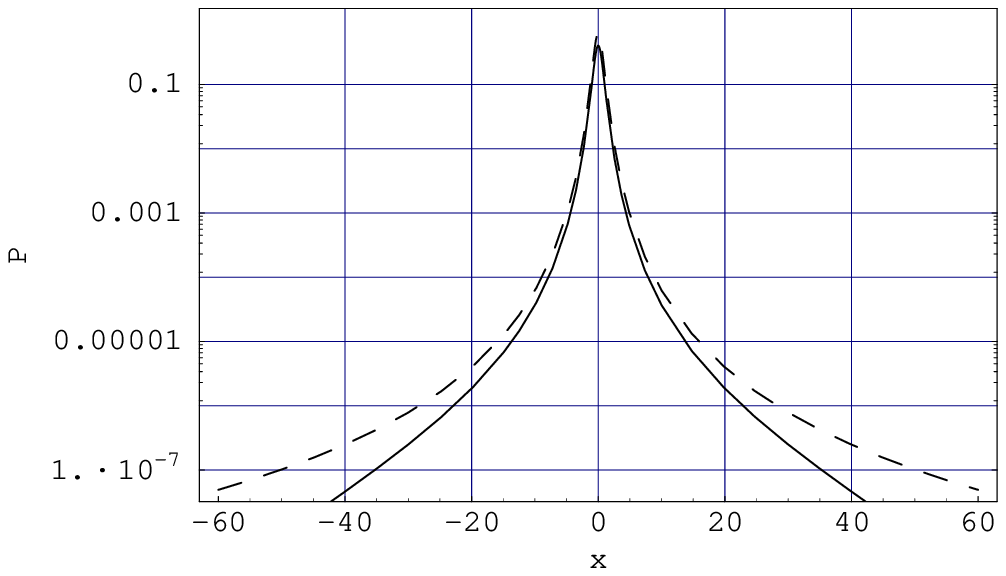}
\includegraphics[width=0.4\textwidth]{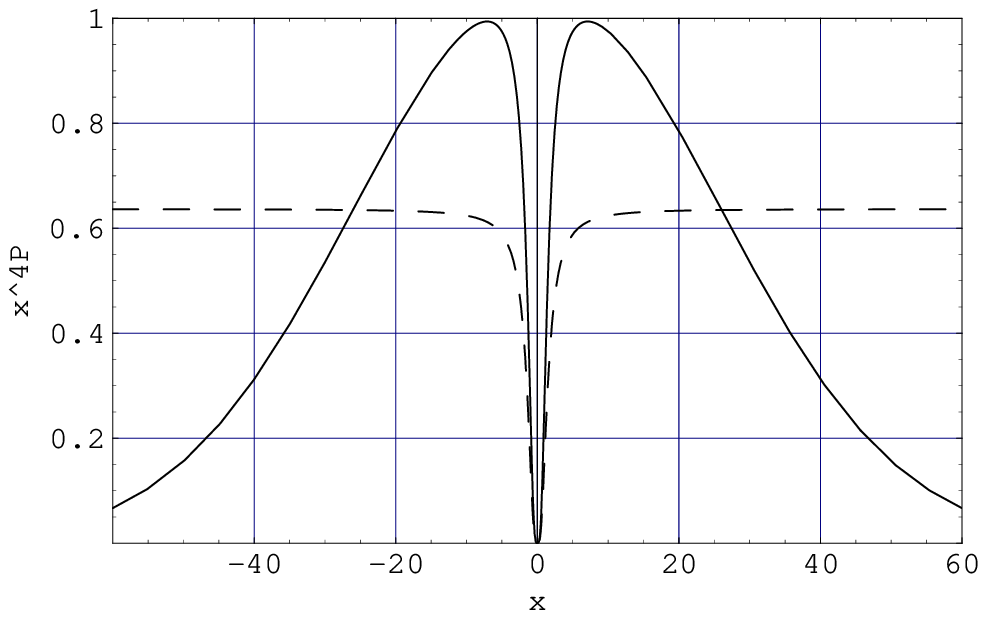}
\end{center}
\caption{ \label{Fig1} Top panel displays probability distribution
functions (\ref{3}) (solid line) and (\ref{2}) (dashed line) of
fluid particle acceleration $P(x)$. Bottom panel displays
$x^4P(x)$ as a function of $x$; solid line corresponds to
Eq.~(\ref{3}); dashed line corresponds to Eq.~(\ref{2}).}
\end{figure}

Crawford, Mordant, and Bodenschatz \cite{Bodenschatz2} reported
new experimental results, which are slightly different from that
reported in earlier experiment \cite{Bodenschatz}, and pointed out
that the Beck's distribution \cite{Beck},
\be\label{2}
P(a) = \frac{C}{(1+\frac{1}{2}\beta(q-1)a^2)^{1/(q-1)}},
\ee
does not correctly capture the tails of the experimental
distribution function. Here, $q=3/2$ (Tsallis entropic index),
$\beta=4$, and $C=2/\pi$ is a normalization constant, all the
parameter values are due the theory. An essential discrepancy is
clearly seen from the contribution $a^4P(a)$ to the fourth moment.

To achieve a better fit, we suggest the modified distribution,
\be\label{3}
P(a) =
\frac{C\exp[-a^2/a_0^2]}{(1+\frac{1}{2}\beta(q-1)a^2)^{1/(q-1)}},
\ee
which is obtained by a "Gaussian screening" of the Beck's
distribution (2). Again, $C$ is a normalization constant,
$\beta=4$, and $q=3/2$, while $a_0$ is a our free parameter, which
we use for a fitting.

Plots of the distribution (\ref{3}) and resulting $a^4P(a)$, for
certain value of $a_0$, are shown in Fig.~1. One can observe a
better fit of the experimental $P(a)$ (top panel, solid line) and
much better agreement of the plot $a^4P(a)$ (bottom panel, solid
line) to the new experimental results presented by Crawford,
Mordant, and Bodenschatz~\cite{Bodenschatz2}, as compared to that
derived from Eq.~(\ref{2}) (dashed line).

We conclude by a few comments.

Despite we use only one parameter, $a_0$, to fit the experimental
data with a good accuracy (values of the other parameters are due
to Eq.~(\ref{2})), it is required to derive Eq.~(\ref{3}) with the
help of the same approach as used by Beck in Ref.~\cite{Beck}, to
provide a self-consistent description. This would allow one to
unravel physical picture lying behind the proposed
phenomenological Gaussian screening in Eq.~(\ref{3}).

We note that the experimentally observed large $a$ asymptotics of
$a^4P(a)$ which is evidently not reproduced by Beck's result may
require some modification of the theoretical set up.

To this end, one may be interested in finding a more appropriate
statistical distribution of the inverse temperature parameter,
instead of gamma distribution, or in modifying of the nonlinear
Langevin equation for velocity, used by Beck.

However, it should be noted that while the gamma distribution is
justified from various aspects, see, e.g.,
Refs.~\cite{Beck,Johal,Aringazin}, as it provides Tsallis
distribution with associated Tsallis entropy, and is known to be
in correspondence with a nonlinear Langevin equation for
fluctuating temperature \cite{Bashkirov} within the framework of
the Landau-Lifschitz theory of fluctuations, the use of some
different distributions \cite{Aringazin,Beck3}, with large
variance of the inverse temperature fluctuations, may send one out
of the established anzatz.


\begin{thebibliography}{4}
\bibitem{Beck}
C. Beck, Phys. Rev. Lett. {\bf 87}, 180601 (2001); {\it
Generalized statistical mechanics and fully developed turbulence},
cond-mat/0110073 (2001) .

\bibitem{Tsallis}
C. Tsallis, J. Stat. Phys. {\bf 52}, 479 (1988).

\bibitem{Bodenschatz}
A. La Porta, G.A. Voth, A.M. Crawford, J. Alexander, and
E.~Bodenschatz, Nature {\bf 409}, 1017 (2001);
 G.A. Voth  {\it et al.}, J. Fluid Mech. {\bf 469}, 121 (2002).

\bibitem{Beck2}
 C. Beck, Physica  A {\bf 277}, 115 (2000);
 Phys. Lett. A {\bf 287}, 240 (2001);
 {\it Non-additivity of Tsallis entropies and fluctuations of temperature},
cond-mat/0105371 (2001).

\bibitem{Wilk}
G. Wilk and Z. Wlodarczyk, Phys. Rev. Lett. {\bf 84}, 2770 (2000).

\bibitem{Bodenschatz2}
A.M. Crawford, N. Mordant, and E. Bodenschatz, {\it Comment on
"Dynamical foundations of nonextensive statistical mechanics"},
physics/0212080 (2002).

\bibitem{Johal}
R. Johal, {\it An interpretation of Tsallis statistics based on
polydispersity}, cond-mat/9909389 (1999).

\bibitem{Aringazin}
A.K. Aringazin and M.I. Mazhitov, {\it Quasicanonical Gibbs
distribution and Tsallis nonextensive statistics},
cond-mat/0204359 (2002).

\bibitem{Bashkirov}
A.G. Bashkirov and A.D. Sukhanov, Zhurnal Eksp. i Theor. Phys.
(Russian JETP) {\bf 95}, 440 (2002).

\bibitem{Beck3}
C. Beck and E.G.D. Cohen, {\it Superstatistics}, cond-mat/0205097
(2002).
\end{thebibliography}
\end{document}